\documentclass[cameraready]{Interspeech}

\title{Supervised Post-training of Speech Foundation Models for Robust Adaptation in Speech Deepfake Detection}

\author[affiliation={1}]{Zihan}{Pan}
\author[affiliation={1}]{Sailor}{Hardik}
\author[affiliation={1},correspondingauthor]{Jinyang}{Wu}

\address{
  $^1$ Institute for Infocomm Research (I2R), Agency for Science, Technology and Research (A*STAR), 1 Fusionopolis Way, 138632, Singapore
}

\email{Pan\_Zihan@a-star.edu.sg, sailor.hardik2000@gmail.com, Wu\_Jinyang@a-star.edu.sg}


\usepackage{comment}
\usepackage{booktabs}
\usepackage{multirow}
\usepackage{threeparttable}
\usepackage{graphicx}
\usepackage{cite}

\begin{document}
\hbadness=10000
\vbadness=10000

\maketitle
\footnotetext[1]{Code available at: https://github.com/pandarialTJU/Mix-Frame-Post-Training.git}
\footnotetext[2]{This research is supported by the National Research Foundation, Prime Minister’s Office, Singapore, and the Ministry of Digital Development and Information, under its Online Trust and Safety (OTS) Research Programme (MDDI-OTS-001). Any opinions, findings and conclusions or recommendations expressed in this material are those of the author(s) and do not reflect the views of National Research Foundation, Prime Minister’s Office,  Singapore, or the Ministry of Digital Development and Information.}

\begin{abstract}
  Large speech foundation models have shown strong potential for speech deepfake detection, but direct fine-tuning is limited by a mismatch between self-supervised pre-training objectives and spoof-specific artifacts. To address this, we propose a mix-frame post-training strategy to create localized spoof-oriented perturbations and use frame-level supervision to encourage the SSL model to learn local inconsistencies that are critical for robust spoof detection. On ASVspoof5, we achieve state-of-the-art EER 4.50\% for a single model without data augmentation. On ASVspoof2021 LA/DF, it further achieves only 0.16\% absolute EER gap between LA and DF, indicating strong and balanced robustness across distinct distortion conditions. These results show that supervised post-training provides an effective and practical way to adapt speech foundation models for robust deepfake detection.
\end{abstract}

\section{Introduction}



Audio deepfakes generated by voice conversion (VC) and text-to-speech (TTS) systems pose increasing risks to biometric authentication, media forensics, and misinformation detection. Early spoofed-speech countermeasures were mainly built on handcrafted front-ends, such as cepstral and phase-related features, coupled with classical back-ends including GMMs and SVMs \cite{wang17_interspeech}. More recently, the field has shifted toward deep neural detectors that learn discriminative cues directly from time--frequency representations or raw waveforms, including LCNN-based systems and graph-attention-based architectures such as AASIST \cite{gomezalanis19_interspeech,jung2022aasist}. Building on this trend, deepfake detectors \cite{liu2024towards,guragain2024speech,pan24c_interspeech,pan2025molex} benefit substantially from self-supervised learning (SSL) speech models such as HuBERT \cite{hsu2021hubert} and WavLM \cite{chen2022wavlm} that provide rich, transferable representations.


However, these encoders are typically pre-trained to model phonetic content and long-range context rather than to amplify the subtle, localized inconsistencies that characterize spoofed speech. Direct fine-tuning starts from representations optimized for the SSL pretext objective (masked prediction/contrastive learning \cite{hsu2021hubert,chen2022wavlm}), which prioritizes phonetic and speaker-related structure rather than spoof-specific artifacts. Consequently, the resulting representation may under-emphasize subtle, local inconsistencies—such as abrupt spectral discontinuities and temporal boundary artifacts—that are critical for robust generalization to unseen VC/TTS attacks \cite{li2024audio,zhang2022partialspoof}. This limitation is further amplified in low-resource anti-spoofing settings, where the labeled target dataset is small or domain-narrow: end-to-end fine-tuning can overfit spurious cues and yield unstable transfer \cite{eom22_interspeech}. In practice, the cues that separate bona fide from spoofed speech are often highly localized and boundary-driven \cite{zhong24_interspeech,tian16_interspeech} (e.g., short segments with vocoder or conversion errors, abrupt temporal/spectral discontinuities, or inconsistent phase \cite{todisco2017constant}), and they may occupy only a small fraction of an utterance. With primarily utterance-level supervision, the fine-tuning signal can be diluted by dominant content information, making the encoder prone to overfitting to known attacks instead of learning general spoof cues. This motivates an intermediate post-training stage \cite{gururangan2020don} that injects local perturbations and provides frame-level supervision, explicitly biasing the SSL feature space toward detecting local inconsistencies.

This paper investigates whether post-training an SSL encoder with frame-label objective can yield representations that are more suitable for deepfake detection. We introduce Mix-Frames Post-Training (MFPT), a perturbation-driven objective that constructs mixed frame sequences to perturb local temporal coherence. The key intuition is that many spoofing artifacts manifest as short-term discontinuities; encouraging the encoder to discriminate or model such perturbations can improve its sensitivity to spoof cues. Our contributions are threefold: (1) the main proposal MFPT, which constructs localized perturbations with frame-level supervision to reshape the representation space before utterance-level fine-tuning; (2) we show that the proposed post-training consistently improves robustness, particularly in low-resource adaptation and out-of-domain evaluation across multiple ASVspoof benchmarks; (3) we provide analysis suggesting that post-training increases sensitivity to local spectral/phase irregularities associated with spoofed speech.


\section{Method}

\subsection{Problem setup}
We consider binary audio deepfake detection with labels $y\in\{0,1\}$ for spoofed and bona fide speech, following common evaluation setups such as ASVspoof 2019 logical access \cite{nautsch2021asvspoof}. Given a waveform $\mathbf{x}\in\mathbb{R}^{T}$ sampled at 16~kHz, an SSL encoder $f_{\theta}$ (e.g., WavLM-Large) produces frame-level representations $\mathbf{h}_{1:N}=f_{\theta}(\mathbf{x})$ with stride $s$ samples per frame, where $T$ and $N=\lfloor T/s \rfloor$ denote the number of raw samples and encoder frames. A downstream detector $g_{\phi}$ maps pooled frame representations to an utterance-level prediction $\hat{y}$.

Our goal is to post-train the $f_{\theta}$ with a frame-level objective that deliberately exposes the encoder to localized discontinuities that resemble deepfake artifacts. Motivated by frame-level mixup/splicing, we propose Mix-Frames: a within-utterance cut-and-paste operation using an utterance from the opposite class, paired with frame-wise supervision.

\begin{figure*}[t]
  \centering
  \includegraphics[width=0.85\linewidth]{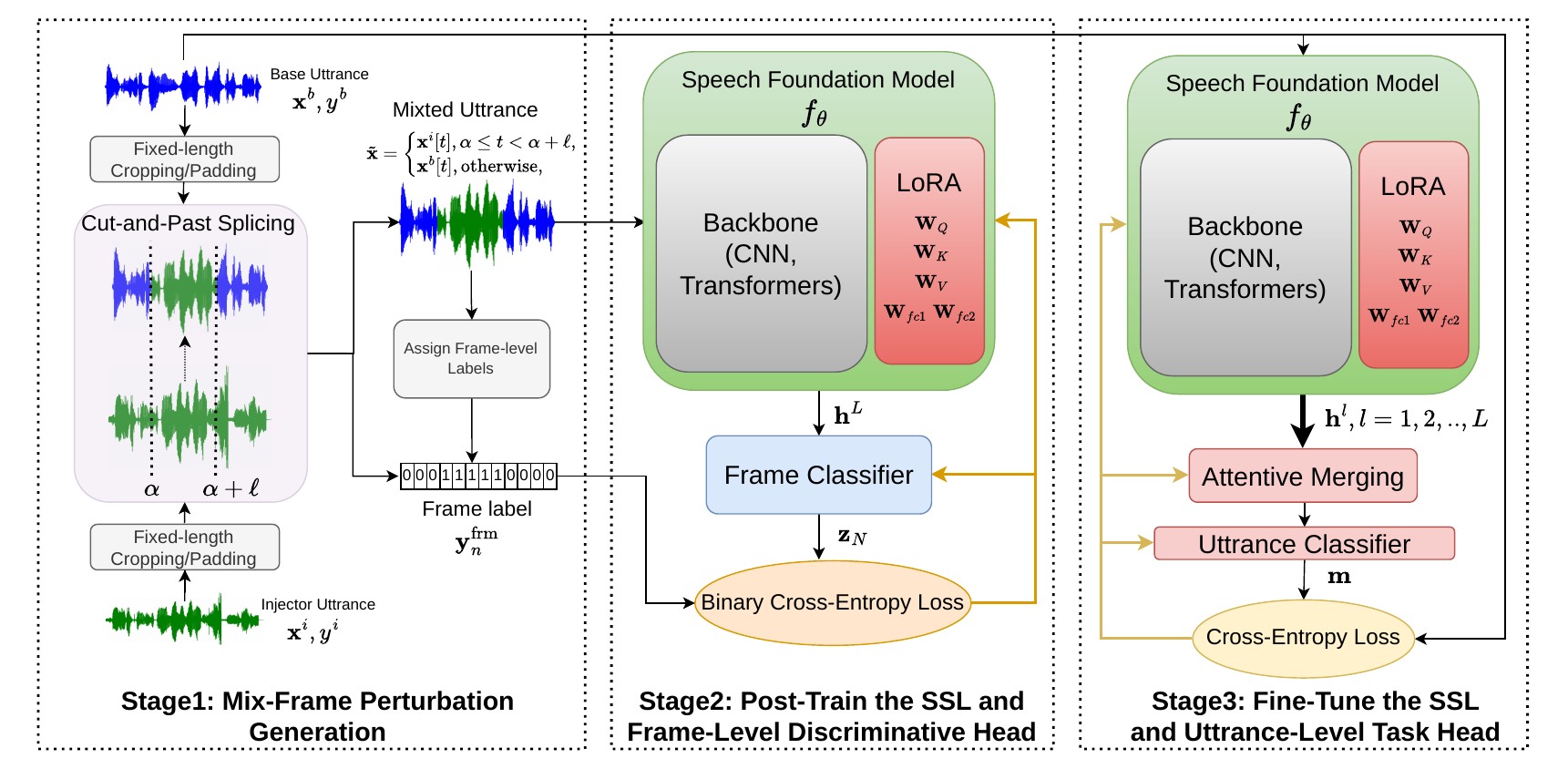}
  \caption{Overview of the proposed framework. The SSL encoder is first post-trained with mix-frames and frame-level supervision, and then fine-tuned for utterance-level deepfake detection.}
  \label{fig:stage_block}
\end{figure*}

\subsection{Stage1: mix-frame perturbation generation}
For each training sample (``Base Uttrance'' in Figure~\ref{fig:stage_block}), we first randomly draw an ``Injector Uttrance'' from the opposite class, ensuring that the mixed segment corresponds to an acoustically plausible region while still introducing an abrupt context change. Concretely, let $(\mathbf{x}^{b}, y^{b})$ denote the base utterance and $(\mathbf{x}^{i}, y^{i})$ the injector utterance, where $y^{i}=1-y^{b}$.

\noindent\textbf{Fixed-length cropping/Padding:}
We randomly crop (or zero-pad) each waveform to a fixed length of $T=64600$ samples to enable efficient batching.

\noindent\textbf{Cut-and-Paste Splicing:}
Then we sample a splice mix ratio $r^{\text{mix}}$ uniformly in a range of $[r^\text{lower}, r^\text{upper}]$, set the splice length $\ell=r^{\text{mix}}*T$, and randomly choose a staringt index $\alpha \sim\mathcal{U}(0, T-\ell)$, $\mathcal{U}$ denoting uniform distribution. The mixed waveform $\tilde{\mathbf{x}}$ is created by replacing the segment in the base waveform with the corresponding segment from the injector:
\begin{equation}
  \tilde{\mathbf{x}}[t]=
  \begin{cases}
    \mathbf{x}^{i}[t], & \alpha \le t < \alpha+\ell,\\
    \mathbf{x}^{b}[t], & \text{otherwise},
  \end{cases}
\end{equation}
This operation introduces two boundary points at $t=\alpha$ and $t=\alpha+\ell$, which encourages the encoder to become sensitive to short-term inconsistencies.

\noindent\textbf{Frame-level labels:}
To provide frame-level supervision for the SSL encoder, we derive the frame labels according to the temporal resolution of WavLM. The SSL encoder commonly operates at a
50~Hz frame rate, each frame corresponds to a stride of $s = 320$ samples for 16~kHz audio. Let the center sample of the $n$-th frame be
\begin{equation}
  c_n = ns + \left\lfloor \frac{s}{2} \right\rfloor, \qquad n=1,\dots,N.
\end{equation}
We then assign each frame label according to whether its center falls inside the injected segment:
\begin{equation}
  y^{\mathrm{frm}}_n =
  \begin{cases}
    y^{i}, & a \le c_n < a+\ell,\\
    y^{b}, & \text{otherwise},
  \end{cases}
  \qquad n=0,\dots,N-1,
\end{equation}
where $y^{b}$ and $y^{i}$ denote the labels of the base and injector utterances, respectively. By jointly constructing the perturbed waveform and the corresponding frame-level labels, it provides a direct supervisory signal for learning spoof-discriminative local inconsistencies during post-training.

\subsection{Stage2: post-training the SSL speech foundation model with a frame-level discriminative head}
We attach a lightweight frame classifier (discriminative head) $q_{\psi}$ on top of the SSL frame representations and optimize the encoder to predict $\mathbf{y}^{\mathrm{frm}}$ from the mixed waveform. In implementation, the encoder returns frame features $\mathbf{h}^{L}_n$ from the $L^{\text{th}}$ SSL layer. The classifier is a linear head that maps each frame feature $\mathbf{h}_n\in\mathbb{R}^{d}$ to a scalar logit:
\begin{equation}
  z_n = q_{\psi}(\mathbf{h}^L_n)=\mathbf{w}^{\top}\mathbf{h}_n + b,
\end{equation}
where $\psi=\{\mathbf{w},b\}$ are the trainable parameters (initialized with Xavier uniform for $\mathbf{w}$ and zeros for $b$). We then use a binary cross-entropy loss with logits in the post-training stage:
\begin{equation}
  \begin{aligned}
    \mathcal{L}_{\mathrm{PT}}(\theta,\psi)
    &= \frac{1}{BN}\sum_{b=1}^{B}\sum_{n=1}^{N}
    \Big[-y^{\mathrm{frm}}_{b,n}\log\sigma(z_{b,n}) \\
    &\quad -\big(1-y^{\mathrm{frm}}_{b,n}\big)\log\big(1-\sigma(z_{b,n})\big)\Big].
  \end{aligned}
\end{equation}
where $y^{\mathrm{frm}}_{b,n}\in\{0,1\}$, $\sigma(\cdot)$ is the sigmoid function. To reduce the number of trainable parameters when adapting SSL speech foundation models, we update the transformer encoder $\theta$ using Low-Rank Adaptation (LoRA \cite{hu2021lora}) adapters inserted into the self-attention projections and dense layers. Specifically, for each transformer layer we apply LoRA to the query, key, and value projections ($W_Q$, $W_K$, $W_V$), and dense layers ($W_{fc1}$, $W_{fc2}$), and optimize only the low-rank updates during both post-training and downstream fine-tuning, while keeping the original backbone weights frozen.

This post-training step encourages speech foundation model $f_{\theta}$ to encode cues that discriminate bona fide versus spoof at a local (frame) level, while mix-frames ensure that such cues appear at varied temporal locations and contexts. After post-training, we discard $q_{\psi}$ and keep the adapted encoder $f_\theta$ for the final fine-tuning stage.

\subsection{Stage3: fine-tuning the SSL speech foundation model with an utterance-level task head}
\label{sec:finetune}

In the final stage of adapting the SSL model to utterance-level deepfake detection, we fine-tune $f_{\theta}$ with the same LoRA parameterization (LoRA on $W_Q$, $W_K$, $W_V$, $W_{fc1}$, $W_{fc2}$) and train the downstream task head using utterance-level labels with cross-entropy loss. Given a training utterance $\mathbf{x}$, we first extract hidden embeddings from the $L$-layer SSL transformer encoder:
\begin{equation}
  \mathbf{H} = \left\{\mathbf{h}^{l}\right\}_{l=1}^{L},
  \qquad \mathbf{h}^{l} \in \mathbb{R}^{B \times N \times D},
\end{equation}
where $B$, $N$, and $D$ denote the batch size, number of frames, and feature dimension, respectively.
The multi-layer representations are then selectively merged layer-wise by the attentive merging module ($\mathrm{AttM}$) \cite{pan24c_interspeech}, where $\mathrm{AttM}$ and the downstream classifier together form the task head for producing utterance-level logits:
\begin{equation}
  \mathbf{m} = \mathrm{Classifier}\left(\mathrm{AttM}\left(\mathbf{H}\right)\right),
  \qquad \mathbf{m} \in \mathbb{R}^{B \times C},
\end{equation}
where $C=2$ corresponds to the bona-fide and spoof classes.

The model is updated with the standard cross-entropy loss:
\begin{align}
  \mathcal{L}_{\mathrm{FT}}
  &= - \frac{1}{B} \sum_{b=1}^{B} \sum_{c=1}^{C} y_{b,c}\log p_{b,c}, \\
  p_{b,c}
  &= \frac{\exp(m_{b,c})}{\sum_{j=1}^{C}\exp(m_{b,j})}.
\end{align}
where $y_{b,c}$ is the one-hot target label of the $b$-th utterance.

Unlike the post-training stage, which imposes frame-level supervision to enhance sensitivity to local spoof artifacts, the fine-tuning stage directly optimizes the model for the final utterance-level bona fide/spoof classification task.

\section{Experiments}

\begin{table}[!ht]
  \centering
  \caption{Effect of the $r^{\text{mix}}$: The ASV5 dataset is used for training and evaluation in this experiment. The classifier is ECAPA-TDNN. Results are reported as EER(\%).}
  \label{tab:exp1_mixratio}
  \setlength{\tabcolsep}{4pt}
  \resizebox{0.85\linewidth}{!}{%
    \begin{tabular}{lccccc}
      \toprule
      Mix ratio $r^\text{mix}$ & 10--30\% & 30--50\% & 50\% & 50--70\% & 70--90\% \\
      \midrule
      EER (\%)  & 4.50     & 5.15     & 5.83 & 7.31     & 5.31 \\
      \bottomrule
    \end{tabular}%
  }
\end{table}

\begin{table}[t]
  \centering
  \caption{Impact of LoRA adapters location and downstream classifier. Results are reported as EER(\%).}
  \label{tab:exp2_lora_location}
  \setlength{\tabcolsep}{4pt}
  \resizebox{0.85\linewidth}{!}{%
    \begin{tabular}{lccccc}
      \toprule

      LoRA & QKV & FFN & \multicolumn{3}{c}{QKV+FFN} \\
      \cmidrule(l{0.2em}r{0.2em}){1-1}\cmidrule(l{0.2em}r{0.2em}){2-2}\cmidrule(l{0.2em}r{0.2em}){3-3}\cmidrule(l{0.2em}r{0.2em}){4-6}
      Classifier & LSTM & LSTM & LSTM & ECAPA & Nes2Net \\
      \cmidrule(l{0.2em}r{0.2em}){1-1}\cmidrule(l{0.2em}r{0.2em}){2-2}\cmidrule(l{0.2em}r{0.2em}){3-3}\cmidrule(l{0.2em}r{0.2em}){4-6}
      Post-Train EER (\%) & 5.56 & 6.01 & 4.68 & 4.50 & 4.55 \\
      w/o Post-Train EER (\%) & 5.64 & 5.05 & 5.74 & 5.18 & 5.14 \\
      \bottomrule
    \end{tabular}%
  }
\end{table}

\begin{table*}[t]
  \centering
  \caption{Low-resource fine-tuning results. The model is post-trained on ASV5 and fine-tuned with various fractions of the ASV19LA training and validation sets. Each entry reports evaluation EER(\%) in the order of ASV19LA, ASV21LA, ASV21DF.}
  \label{tab:exp3_lowresource}
  \setlength{\tabcolsep}{5pt}
  \resizebox{0.80\textwidth}{!}{%
    \begin{tabular}{lccccc}
      \toprule
      Train fraction of ASV19LA & 20\% & 40\% & 60\% & 80\% & 100\% \\
      \midrule
      Post-Train &
      (1.37, 7.47, 4.34) &
      (0.89, 6.04, 6.73) &
      (0.72, 4.84, 4.11) &
      (0.59, 4.46, 4.10) &
      (0.44, 3.88, 4.04) \\
      w/o Post-Train &
      (1.47, 7.00, 9.32) &
      (1.19, 7.18, 7.02) &
      (0.84, 7.36, 6.77) &
      (0.76, 6.57, 7.35) &
      (0.58, 7.10, 6.79) \\
      \bottomrule
    \end{tabular}%
  }
\end{table*}

\begin{table*}[t]
  \centering
  \caption{Comparison with representative systems on ASVspoof benchmarks.
    EER(\%) is reported on the evaluation sets as stated in the cited papers.
  For ASVspoof2021, we additionally report the average EER across LA and DF, the worst-case EER $max(\text{EER}^{21LA}, \text{EER}^{21DF})$, and the absolute gap between the two conditions $|\text{EER}^{21LA}-\text{EER}^{21DF}|$.}
  \label{tab:sota_comparison_v2}
  \setlength{\tabcolsep}{5.5pt}
  \renewcommand{\arraystretch}{1.12}

  \begin{threeparttable}
    \resizebox{0.80\textwidth}{!}{
      \begin{tabular}{c l c c c c c c c c}
        \toprule
        \multirow{2}{*}{\textbf{Benchmark}} & \multirow{2}{*}{\textbf{Method}} & \multirow{2}{*}{\textbf{Train}} & \multirow{2}{*}{\textbf{Aug}} & \multirow{2}{*}{\textbf{Fusion}}
        & \multicolumn{5}{c}{\textbf{ASVspoof21 eval set EER(\%)}} \\
        \cmidrule(lr){6-10}
        & & & & & \textbf{ASV21LA} & \textbf{ASV21DF} & \textbf{Avg.} & \textbf{Worst} & \textbf{Gap} \\
        \midrule

        \multirow{5}{*}{\rotatebox[origin=c]{90}{
            \begin{tabular}{c}ASVspoof\\21LA/21DF
        \end{tabular}}}
        & Wav2vec2-XLSR \cite{wang22_odyssey}         & 19LA & Yes & Single & 7.18 & 5.44 & 6.31 & 7.18 & 1.74 \\
        & Do\~nas et al. \cite{martin2022vicomtech}        & 19LA & Yes & Single & 3.54 & 4.98 & 4.26 & 4.98 & 1.44 \\
        & WavLM+MFA \cite{guo2024audio}             & 19LA & Yes & Single & 5.08 & 2.56 & 3.82 & 5.08 & 2.52 \\
        & WavLM+ASP (SCL+CE) \cite{tran24_interspeech} & 19LA & No  & Single & 3.31 & 4.47 & 3.89 & 4.47 & 1.16 \\
        \cmidrule(lr){2-10}
        & \textbf{Ours}                                & 19LA & No & Single & 3.88 & 4.04 & 3.96 & \textbf{4.04} & \textbf{0.16} \\
        \midrule

        \textbf{Benchmark} & \textbf{Method} & \textbf{Train} & \textbf{Aug} & \textbf{Fusion}
        & \multicolumn{5}{c}{\textbf{ASVspoof5 Track 1 (eval) EER(\%)}} \\
        \midrule

        \multirow{7}{*}{\rotatebox[origin=c]{90}{ASVspoof5}}
        & SZU-AFS \cite{xu24_asvspoof}                    & 5 & Yes & Fusion & \multicolumn{5}{c}{4.04} \\
        & SLIM (Reality Defender) \cite{zhu24_asvspoof}  & 5 & Yes & Single & \multicolumn{5}{c}{5.50} \\
        & WavLM-ResNet18-SA fusion \cite{chan24_asvspoof}& 5 & Yes & Fusion & \multicolumn{5}{c}{7.01} \\
        & BUT system (fused) \cite{rohdin24_asvspoof}    & 5 & No  & Fusion & \multicolumn{5}{c}{9.28} \\
        & SSL-IVSPT \cite{guo24_asvspoof}                & 5 & Yes & Single & \multicolumn{5}{c}{5.99} \\
        & MoLEx \cite{pan2025molex}                      & 5 & No & Single & \multicolumn{5}{c}{5.56} \\
        \cmidrule(lr){2-10}
        & \textbf{Ours}                                  & 5 & No  & Single & \multicolumn{5}{c}{\textbf{4.50}} \\
        \bottomrule
      \end{tabular}
    }
  \end{threeparttable}
\end{table*}

\subsection{Datasets and protocols}
We evaluate on several public benchmarks covering both classical spoofing and modern deepfake conditions: ASVspoof 2019 LA (ASV19LA), ASVspoof 2021 LA and DeepFake (ASV21LA, ASV21DF) \cite{liu2023asvspoof}, ASVspoof 5 (ASV5) \cite{wang24_asvspoof}. For each benchmark, we follow the official train/dev/eval splits and report results on the evaluation partition unless stated otherwise.

\subsection{Model configuration}
We use WavLM\_Large as the backbone encoder $f_{\theta}$ in this work. The input waveform is resampled to 16~kHz and truncated or padded to  fixed 4s duration when required by the training recipe. In both post-training and fine-tuning, we update $f_{\theta}$ using LoRA adapters (rank=32) on the self-attention projections (Q,K,V) and dense layers (FFN). Unless otherwise specified, all original WavLM\_Large parameters remain frozen. Specifically in fine-tuning, layer-wise SSL embeddings are aggregated into one representation using the attentive merging module \cite{pan24c_interspeech}. On top of the embedding, we compare three classifier back-ends: (i) a BiLSTM-based classifier (same used in \cite{pan2025molex}) (ii) ECAPA-TDNN \cite{desplanques2020ecapa}, and (iii) Nes2Net \cite{Nes2Net}. All back-ends are trained with the same utterance-level cross-entropy objective. Both post-training and fine-tuning are performed with DDP on 4 NVIDIA H200 GPUs with a batch size of 256; the learning rates were set to $4\times10^{-4}$ for post-training and $5\times10^{-5}$ for fine-tuning. The experiment results are reported in EER.

\subsection{Results}


We first investigate the effect of the injector mix ratio $r^{\text{mix}}$ in the post-training stage. As shown in Table~\ref{tab:exp1_mixratio}, the best performance is achieved with a relatively low mixing range of 10--30\%, yielding an EER of 4.50\%. In general, larger mixing ratios lead to worse performance, with the largest degradation observed at 50--70\% (7.31\% EER). This suggests that moderate mixing is sufficient to introduce localized spoof-relevant perturbations, whereas overly strong mixing may distort the original speech structure and reduce the effectiveness of the proposed post-training strategy. Although the performance at 70--90\% (5.31\%) partially recovers, it remains inferior to the low-ratio setting. Therefore, we adopt 10--30\% as the default configuration in the subsequent experiments.

Next, we ablate various LoRA adapter placements (multi-head attention projection (QKV), dense layers (FNN), or both) and downstream utterance-level classifiers, with and without supervised post-training. As shown in Table~\ref{tab:exp2_lora_location}, post-training improves most configurations, indicating a better initialization for subsequent fine-tuning. The best result is achieved by adapting both QKV and FFN together with the ECAPA classifier, yielding an EER of 4.50\%. In contrast, restricting LoRA to only QKV or only FFN is less effective, especially after post-training (5.56\% and 6.01\%, respectively). These results suggest that post-training is better supported by coordinated adaptation of both attention and FFN submodules, while a stronger utterance-level classifier can better exploit the refined representations.

Table~\ref{tab:exp3_lowresource} shows the results of low-resource fine-tuning, where the model is first post-trained on ASV5 and then fine-tuned using different fractions of the ASV19 LA training set. Evaluation is conducted on both the in-domain ASV19LA benchmark and the out-of-domain ASV21LA/DF benchmarks. A clear trend is observed: under the post-training setting, performance improves progressively as the amount of target-domain training data increases, and consistently outperforms the counterpart without post-training. The advantage is particularly pronounced under limited-data conditions (e.g., 20\%: 9.32\% $\rightarrow$ 4.34\% on ASV21DF), while the gains on ASV19LA are smaller but remain consistent, suggesting that direct fine-tuning is already effective in the matched in-domain setting. Importantly, the strongest gains are observed in the most challenging regime, i.e., when target-domain data are scarce and evaluation is performed under out-of-domain conditions. This indicates that the post-training stage learns a more transferable spoof-aware representation, which is especially valuable for practical deepfake detection where labeled target data are often limited and test conditions may differ substantially from the fine-tuning data domain.


Finally, Table~\ref{tab:sota_comparison_v2} summarizes representative benchmark results on ASVspoof5 and ASVspoof2021 LA/DF. On ASVspoof5, our method achieves an EER of 4.50\%, which is SOTA for a single model without data augmentation. This result is particularly encouraging given that several competing systems rely on augmentation or fusion to further boost performance. On ASVspoof2021, although our method does not always attain the lowest EER on each individual benchmark, it still achieves SOTA-level performance for a single model without augmentation, while showing much more consistent behavior across LA and DF than competing approaches. In particular, compared with several augmentation-based systems, our method remains competitive in average EER, while achieving the best worst-case EER and by far the smallest LA--DF gap (0.16), indicating substantially better cross-condition (21LA/DF) stability. This result is especially meaningful because ASVspoof2021 LA and DF represent two different forms of domain shift: LA mainly reflects coding/compression and transmission effects, whereas DF focuses on compressed manipulated speech distributed online via social-media codecs \cite{liu2023asvspoof}. From this perspective, the advantage of our method is not merely strong performance on a single benchmark, but its ability to maintain stable generalization across distinct distortion conditions. This suggests that the proposed supervised post-training stage encourages the model to learn more transferable spoof-discriminative cues, rather than overly adapting to a particular benchmark-specific artifact pattern. For practical deepfake detection systems, such balanced robustness is often more desirable than obtaining a lower EER on one condition at the cost of degraded performance consistency on others.

\section{Conclusion}
\label{sec:conclusion}

This work proposed a supervised post-training framework for speech foundation models for robust speech deepfake detection. By introducing an intermediate post-training stage with mix-frame perturbation generation and frame-level supervision, the proposed method explicitly biases the learned representation toward localized spoof artifacts before final utterance-level fine-tuning. Results show that this strategy consistently improves robustness, particularly in low-resource and out-of-domain settings. In addition, our method achieves SOTA performance for a single model without data augmentation on ASV5, while maintaining highly balanced performance on ASV21LA/DF. These findings suggest that supervised post-training can serve as a promising adaptation paradigm for speech foundation models, with the potential to improve robustness and transferability for deepfake detection under diverse real-world conditions.


\section{Generative AI Use Disclosure}
AI tools were employed to enhance the clarity and flow of the text in this manuscript. The authors take complete responsibility for the integrity and originality of the final paper.

\bibliographystyle{IEEEtran}
\bibliography{mybib}

\end{document}